# Nanoengineering of Anisotropic Materials for Creating the Active Optical Cells with Increased Energy Efficiency

Nazariy Andrushchak
Department of Computer-Aided Design Systems
Lviv Polytechnic National University
Lviv, Ukraine
nandrush@gmail.com

Petra Goering
Department of Research and Development
SmartMembranes GmbH
petra.goering@smartmembranes.de

Anatoliy Andrushchak
Department of Telecommunication
Lviv Polytechnic National University
Lviv, Ukraine
anat@polynet.lviv.ua

*Abstract* — In this paper the state-of-the-art for exploiting the unique physical and chemical properties of crystalline materials and their possible applications for development of crystalline nanocomposites with the tailored anisotropy were discussed. Using a method of growing crystals from a solution the KDP and TGS nanocrystalites were grown in the mezopores of the $Al_2O_3$ matrix. To investigate the developed nanocrystalline samples the X-ray analysis was used. A form of the obtained diffractograms and our calculations show the predominant crystallization of the KDP and TGS in the direction [100], which coincides with the direction of the cylindrical pores. Thus, diffractograms contain only reflexes of crystallographic planes (200) and (400). The achieved result indicates the possibility of growing of nanocrystals in mesopores of such $Al_2O_3$ structures developed by us and provides the prospect of creating optical cells with increased energy efficiency.

## I. INTRODUCTION

Energy efficient systems that allow light manipulation, for example, modulation of laser beams, by external controlling actions are crucial for further development of optoelectronics, laser devices industry and important for many other areas. The input signal in such systems is typically represented by electromagnetic wave which can be controlled by electric [1] or magnetic [2] field, mechanical pressure [3] and surface or bulk acoustic wave [4]. The proposed idea in this research topic focuses specifically on the case, when electromagnetic wave is in optical (visible or infrared) and quasi-optical (sub-terahertz) spectral ranges. There are interaction effects that cause a link between the controlling factor and the input signal which are described using tensors of high orders [5]. The performance of a device based on these effects improves when the magnitude of such interactions increases. Conventional and straightforward way of achieving increased response is to use stronger controlling field. The result evidently depends on the properties of the medium, where the interaction occurs and on the interaction geometry, i.e. mutual orientations of the propagation direction and polarization of the optical signal and the direction (and, in case of the acoustic field, the polarization) of the external field. If anisotropic crystalline medium is considered as a working material, the signal and the field orientations with respect to the principal or symmetry axes in a crystal play an important role

Determination of the precise interaction geometry that would be optimal in the sense of efficiency is an innovative way towards increasing the device performance. There are a number of aspects to this: while keeping the sensitivity of the device constant, one could decrease its weight, external dimensions and power consumption. In the case of anisotropic media, optimization of the interaction geometry is extremely challenging. It requires simultaneous consideration of a number of physical effects involved, e.g. electro-optic and inverse piezo-electric, or elasto-optic and elastic effects when analyzing respectively electro- (in acentric crystals) or acousto-optic (in all media) interactions [6-8]. Our previous research indicates that the appropriate efficiency increase in the case of anisotropic materials can as huge as orders of magnitude and, in general, the most efficient interaction geometry of these field-induced effects does not correspond to the principal (crystallographic) axes - an important fundamental conclusion, made by the authors in [9].

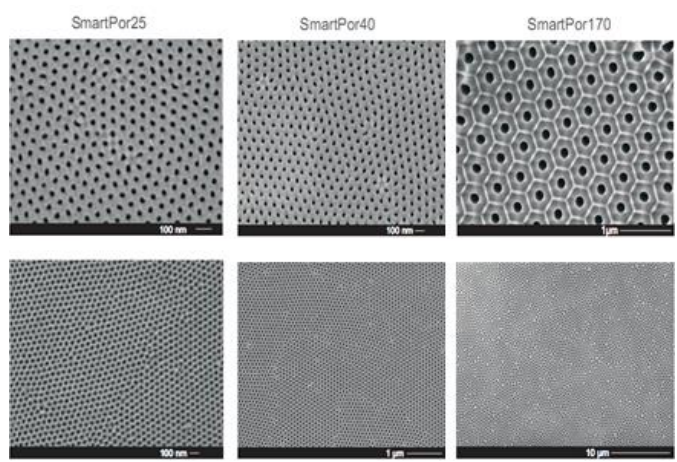

Fig. 1. Different variants of nanoporous membrans manufactured by the SmartMembrans company

All of the above applies directly to bulk crystalline materials in which anisotropy occurs naturally. On the other hand, the development of nanotechnologies has already stimulated extensive investigations of nanocrystalline materials or crystalline nanocomposites in the form of thin plates or membranes (see Fig. 1).

Their characteristics, particularly the anisotropy [10] of their physical properties, are promising for many optical sensing applications [11]. It is known that host mesoporous membrans or matrices (i.e. host materials containing pores ranging in diameter roughly between few nanometers and few tens of nanometers) exhibit anisotropic properties and behave as optically uniaxial crystals [12]. If the pores of a mesoporous material, which is considered as a host matrix, are additionally filled with another anisotropic material, then such novel crystalline nanocomposites will acquire more clearly pronounced anisotropy [13] and could reveal unique properties. However, theoretical description of anisotropic porous material involves complicated mathematical representation [14]. Thus, the new type of developed crystalline nanocomposite can be used as a sensitive element for producing electro-optical cells (see Fig. 2)

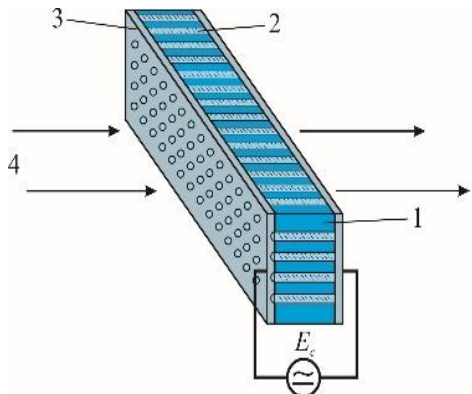

Fig. 2. An illustrating of crystalline nanocomposite as a sensitive element for electro-optic application (1 – nanoporous matrix; 2 – nanocrystallites grown in preferable direction; 3 – transparent electrodes; 4 – laser light)

In relation to the "state of art", several successful attempts to synthesize such composite structures have already been reported, e.g. those involving KDP ($KH_2PO_4$) [15], $LiNbO_3$ [16], triglycine sulphate and Rochelle salt [17] as nanofillers. For example, the authors in [14] have filled the pores with a saturated liquid KDP solution, where the process becomes possible due to capillary forces. After drying, KDP nanocrystals have been formed inside the pores.

Such nanocomposites have already proven their tremendous potential in nonlinear optics and the electro-optics [18]. For instance, it is known that the quadratic nonlinear optical and electro-optic properties can be efficiently modified by controlling the orientation of organic nanocrystals [19]. Furthermore, many promising possibilities arise if one applies the pore engineering techniques. Precise methods of pore engineering and functionalization open new possibilities for adjusting the properties of the host materials [12]. There are reports on how to manipulate crystal orientation in cylindrically shaped nanosized pores [20]. In addition to abovementioned, successful synthesis of nanoparticles of an important optical material, $LiNbO_3$, in mesoporous matrix has been reported [16]. Moreover, the authors in [21] have embedded lithium niobate microtubes inside macroporous silicon to achieve superior optical properties.

Thus, the main goal of the research presented in this paper is to develop original approach to enhancing the energy efficiency of the induced optical interactions in both intrinsically and artificially anisotropic materials at macroscale and nanoscale and to implement the results into a relevant innovative product. The key technology for the realization of this goal is based on innovative optimization approaches, the main features of which are described on our web-site: http://tk-lab.lp.edu.ua/projects/eu-7fp.html. This technology will be used to design active elements of optoelectronic devices for the optical and quasi-optical (sub-terahertz) ranges. The term "anisotropic materials" here represents both bulk crystals and novel crystalline nanocomposites with tailored anisotropy or liquid crystal-based nanocomposites (hereafter referred to as nanocrystalline materials, according to description above).

II. EXPERIMENT

To create the above-described nanostructures filled with crystalline materials, it is proposed to use nanoporous matrices produced by the German company SmartMembrane Ltd. made of 0.1-0.2 mm thick aluminum oxide ($Al_2O_3$) and with a diameter of nanopores in the range from 25 to 80 nm. A general view of a nanoporous matrix that is made using a process of electrochemical etching of pure aluminum is depicted in Fig. 1. The product of this production is an amorphous oxide, which can be changed to a crystalline structure by heating. The pore diameter varies in the range of nanometers. The membranes in Fig. 1 has round pores of unique diameter, with an error of less than 10%, which in turn remains the same along the entire depth of the pores.

Using a method of growing crystals from a solution, which is based on a slow controlled lowering of the temperature of the solution from a supersaturated aqueous solution of potassium dihydrogen phosphate (KDP or $KH_2PO_4$) as well as triglycine sulfate $(NH_2CH_2COOH)_3H_2SO_4$ (TGS), we have grown such nanocrystallites in mesopores of the $Al_2O_3$ matrix [22]. As nanoporous matrices 0.1 mm thick membranes were used. The diameter of the pore in one sample was 35 nm, in another sample - 50 nm. The pores in the form of solid hollow cylinders were orthogonal to the plane surface of the samples and were distributed statistically uniformly (see Fig. 1).

X-ray analysis was used to study flat amorphous $Al_2O_3$ samples of 0.1 mm thickness with the inclusion of KDP and TGS nanocrystallites in the pores of the investigated matrix. As an example, the experimental data of X-ray diffraction analysis for a sample with a pore diameter of 35 nm and KDP nanocrystallites are shown in Fig. 3.

X-ray analysis was performed by powder diffractometry using CuKα irradiation in the mode U = 32 kB, I = 12 mA. The half-width β of the diffraction reflections was measured for the CuKα1 radiation component with a wavelength of λ = 1.540562Å.

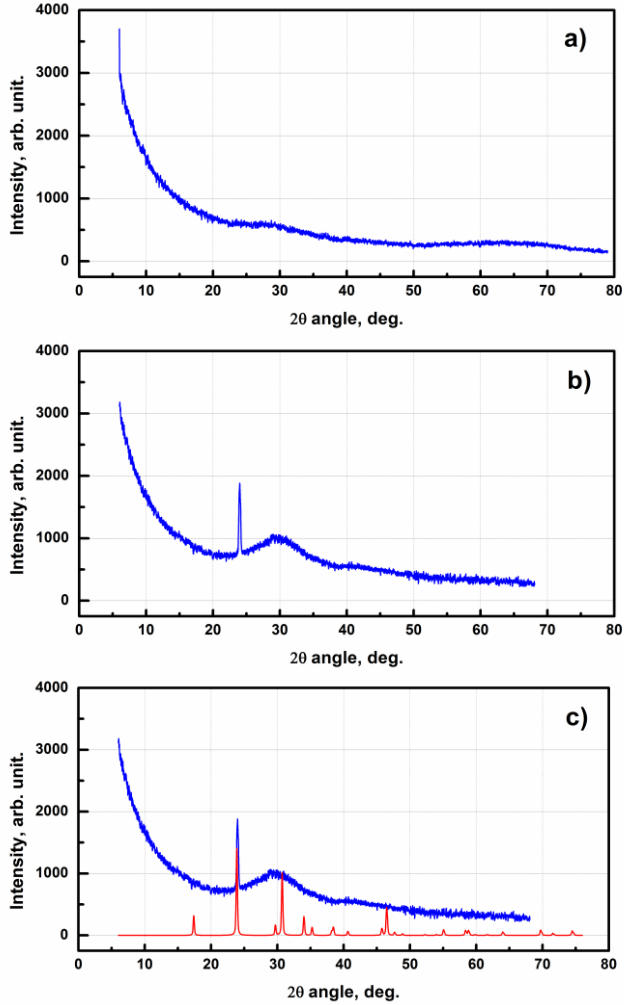

Fig. 3. Experimental data of X-ray diffraction analysis for a) the $Al_2O_3$ matrix without impurities, b) nanoporous matrix filled with KDP crystals in pores, c) comparison of the diffraction curve for a nanoporous matrix filled with KDP crystals in pores (blue curve) and a spectra of a pure KDP crystal (red curve)

The diffraction patterns of the $Al_2O_3$ matrix of sample 1 without the presence of KDP were obtained (see Fig. 3a). Moreover, for this sample two diffraction reflections from single crystal silicon sample with the cut along the (111) and (110) planes were achieved. Reflections of radiation from these silicon surfaces are in the vicinity of Bragg angles 2Θ for reflections (200) and (400) KDP (see Fig. 4). Thus, a form of the obtained diffractograms and our calculations show the predominant crystallization of the KDP in the direction [100], which coincides with the direction of the cylindrical pores. Thus, diffractograms contain only reflexes of crystallographic planes (200) and (400), where orthogonal to these planes crystallization from crystallites into a solid phase occurs. This circumstance and the dislocation structure of single-crystal silicon make it possible to apply these reflexes to subtract the instrumental component from the (200) and (400) KDP reflections in the calculations of the microstructure of the $KH_2PO_4$ phase crystallites.

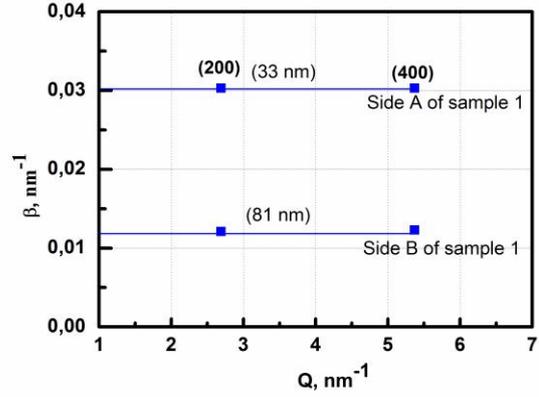

Fig. 4. Reflections of radiation from the surface of sample 1 (diameter of pore 35 nm)

Measurements of the half-width of reflections are preceded by subtraction operations of the CuKα2 constituent of the doublet Kα line and the smoothing operation. An example of such a transformation for the reflex (200) is shown in Fig. 4.

Microstructure of KDP crystallites (block sizes $D$ and distortion of the crystal lattice $\varepsilon$) was determined using the Williamson-Hall method. On the basis of the proposed equation, a graph of linear dependence is constructed as

$$\beta = \frac{1}{D} + 2\varepsilon Q, \quad (1)$$

in coordinates $\beta$ [nm$^{-1}$] and $Q$ [nm$^{-1}$], which was carried out for two reflexes from planes (200) and (400):

In Table 1 a main parameters for calculating the block size $D$ of nanocrystallites KDP for side B of sample 1 with pore diameters of 35 nm are presented.

TABLE I. CALCULATED VALUES FOR SIDE B OF SAMPLES 1

| | hkl | 2Θ, deg | Q, [nm$^{-1}$] | $\beta_s$, deg | $\beta_s \times \cos\Theta$ | D, nm |
|---|---|---|---|---|---|---|
| **Side B** | (200) | 24,0 | 2,69 | 0,11 | 0,00188 | 81 |
| | (400) | 48,8 | 5,37 | 0,12 | 0,00191 | |

III. CONCLUSIONS

Thus, for the first time, using the X-ray diffraction analysis method, the growth effect of the KDP nanocrystallites in the mesopores of the nanoporous $Al_2O_3$ matrix was discovered and confirmed. The crystals found in pores of diameter 35 nm and 50 nm were grown in the direction [100], which coincides with the direction of production of these cylindrical pores. Thus, the achieved result indicates the possibility of growing of nanocrystals in mesopores of such $Al_2O_3$ structures developed by us and provides the prospect of creating active cells with increased energy efficiency. The latter can be realized on the basis of nanodynamics of anisotropic materials using our original approach to finding the optimal geometry of the application of such materials and the creation of crystalline nanocomposites with tailored anisotropy.

ACKNOWLEDGMENT

This result of investigation is a part of a project that has received funding from the European Union's Horizon 2020

research and innovation programme under the Marie Skłodowska-Curie grant agreement No 778156, as well as was supported by Ministry of Education and Science of Ukraine in the frames of the projects entitled "Anisotropy" (registration # 0116U004136) and "Nanocomposite" (registration # 0116U004412).